\documentclass[twoside,twocolumn]{article}

\usepackage{setspace} 

\usepackage{physics}

\usepackage{blindtext} 

\usepackage[sc]{mathpazo} 
\usepackage[T1]{fontenc} 
\usepackage{microtype} 

\usepackage[english]{babel} 

\usepackage[hmarginratio=1:1,top=32mm,columnsep=20pt]{geometry} 
\usepackage[hang, small,labelfont=bf,up,textfont=it,up]{caption} 
\usepackage{booktabs} 

\usepackage{lettrine} 

\usepackage{enumitem} 
\setlist[itemize]{noitemsep} 

\usepackage{abstract} 

\usepackage{titlesec} 
\renewcommand\thesection{\arabic{section}} 
\renewcommand\thesubsection{\Alph{subsection}} 
\titleformat{\section}[block]{\large\scshape\centering}{\thesection.}{1em}{} 
\titleformat{\subsection}[block]{\normalsize\itshape\centering}{\thesubsection.}{1em}{} 

\usepackage{fancyhdr} 
\usepackage{titling} 

\usepackage{hyperref} 

\usepackage{amsmath}
\usepackage{amsfonts}
\usepackage{amssymb}

\usepackage{textcomp}

\pretitle{\begin{center}\huge\bfseries} 
\posttitle{\end{center}} 
\title{{The Action Principle in Market Mechanics}} 
\author{%
\textsc{J. T. Manhire} \\ 
\normalsize \textit{Texas A\&M University, College Station, Texas}
}
\date{} 
\renewcommand{\maketitlehookd}{%
\begin{abstract}
\noindent This paper explores the possibility that asset prices, especially those traded in large volume on public exchanges, might comply with specific physical laws of motion and probability. The paper first examines the basic dynamics of price displacements in financial markets and finds one can model this dynamic as a harmonic oscillator at local ``slices'' of elapsed time via a homogeneous coordinate system. Price and time coordinates here are Wigner invariant meeting the threshold of mathematical law discovery in markets. Based on this finding, the paper theorizes that price displacements are constrained, meaning they have extreme values beyond which they cannot go for defined periods. By showing that price displacements also comply with the principle of stationary action, the paper explores a method for measuring specific probabilities of future price displacements based on prior historical data. Testing this theory with limited markets suggests it can make forecasts consistent with historical data as to constraints on extreme price movements during market ``crashes'' and probabilities of specific price displacements at other times.
\end{abstract}
}

\begin{document}

\maketitle


\section{Introduction}

\lettrine[nindent=0em,lines=3]{T}he goal of this paper is to share a new way of looking at financial markets as physical systems. It explores the possibility that asset prices for financial markets, especially those traded in large volume on public exchanges, might comply with specific (even if analogous) physical laws of motion given a specific coordinate perspective of elapsed time, displaced price, and displaced phase in the complex plane.

In 1991, Landauer proposed that information is physical, especially information that is fixed in some tangible medium \cite{landauer}. Such media can include ink on paper, impressions in clay tablets, or even organized binary code in electronics. This leaves one to ponder, ``If the information fixed in a tangible medium is physical, to what extent does that information comply with physical laws?'' 

Looking at something as basic as a standard stock chart, one can see how the information about market movements is represented in such a way as to be analogous to a graph of motion over time. To this end, the paper seeks to apply knowledge and techniques found in theoretical physics specifically to financial markets to test the consequences of Landauer's proposition in this area. 

This is by no means the first such attempt to analyze economic systems with tools of the physical sciences. Many economists in the early 20th century borrowed almost term for term the classical physics of the 1880s \cite{mirowski}. Others have written specifically on financial markets through the lens of physics in the late 20th century \cite{osbourne}, and the econophysics school continues examining financial markets with the tools of statistical mechanics \cite{gontis}. Still, this paper claims to be the first to imagine the physical system itself in this particular way.\footnote{The author is sensitive to the general rebuke that ``economics is not physics.'' \cite{mises} Yet, the goal here is not philosophical; it is simply to see where Landauer's proclamation leads if one takes it literally and follows it to its logical conclusions. If those conclusions end up being consistent with historical market data then perhaps there is more here for others to investigate and philosophical questions with which others can grapple.}

An initial postulate follows from an obvious implication of the paper's objective:

\textit{I. By analogy, financial markets comply with, even if they are not necessarily subject to, certain physical laws.} \\
If one does not assume this, the entire enterprise would be without logical motivation.

Mantegna and Stanley \cite{mantegna} define a financial market as a system fueled by a large number of interactions between buyers and sellers in perpetual search for an appropriate price for an asset. Broken down further, this definition explains that a financial market is a system, the mechanics of which relate solely to changes in price and the sole drivers of these mechanics are the interactions of buyers and sellers. The remaining two postulates follow from this definition:

\textit{II. Observed market mechanics are the result of local systems that can be defined by their respective changes in price and time.} 

\textit{III. The only phenomenon that directly affects an asset's price displacement is the net effect generated by the trading activities of buyers and sellers at each instance of elapsed time.}

From these postulates, the paper attempts to follow logical conclusions to a theoretical end. As shared in the final section, the theoretical results from these assumptions are consistent with historical data---at least for the few assets against which the theory has been tested. While every attempt is made throughout this paper to clearly convey the process that leads from observation to theory, there is not attempt at true mathematical rigor.

The paper is presented in seven sections, the first being this introduction. Section 2 explores the basic dynamics of an asset's price displacement at each instance in time. Section 3 investigates the forces that drive this price displacement. Section 4 explores the application of the action principle to the asset's movement through the price dimension.  Section 5 derives probability measures from the dynamics discussed.  Section 6 attempts to test the theory by comparing results obtained from traditional analyses with those obtained solely from the theoretical model. Initial comparisons with historical asset price data appear promising. Section 7 closes the paper with a brief summary and recommendations for future research.


\section{Local Systems} \label{Postulate I}

We begin with first principles by observing financial market data fixed in a tangible medium and asking, ``What can we deduce about the most fundamental aspects of a market's change in price over some elapsed period?'' The answer seems clear. The price of an asset---the object of the system we call a financial market---goes up, down, or stays the same when comparing the asset's final and initial prices over some period. Seen this way, it's reasonable to talk about the asset as moving \textit{through} price over time.\footnote{We can also talk about the price moving \textit{around} the asset if we wish. After all, does the ship move through the water or does the water move around the ship? In the end, the chosen reference frame seems more intuitive.}

Imagine a single dimension of an inertial frame of reference for an asset. This is not a spatial dimension in the traditional sense, but rather a dimension of price. An asset moves through this price dimension consistent with our third postulate.

Let's explore the basic dynamics of an asset's movement through this price dimension, that is, its \textit{price displacement}. Displacement involves moving something from one position to another over some elapsed period. Therefore, to arrive at the most basic dynamic of an asset's price displacement we must first try to understand the most basic instance of elapsed time.

\subsection{Elapsed Time}

Assume some elapsed period $t$, where $t$ denotes a change in time from $t_A$ to $t_B$, or $t :=t_B-t_A$. If we divide this period by some integer $n$ we get $n$-number of time slices, each of which we'll call $\tau$. If $n$ is very large then each time slice is very small. These slices need not be precisely equal intervals, but for our purposes we will assume they are.

We can identify each time slice according to an index $\epsilon$ that runs from the states $A$ to $B$ such that 
\begin{displaymath}
\epsilon \in \{A=1<2<\cdots<n-1<n=B\}. 
\end{displaymath}
The identifier $\tau_\epsilon$ indicates the time slice that begins at the state $\epsilon$; that is, $\tau_A=\tau_1=t_1 \rightarrow t_2$, $\tau_2=t_2 \rightarrow t_3$, $\tau_3=t_3 \rightarrow t_4$, etc., until we finally get $\tau_{n-1}=t_{n-1} \rightarrow t_{n}=t_B.$ 

The slice $\tau$ should not be confused with the notion of $dt$. The former is a period of elapsed time that can be arbitrarily small but never be zero. The latter is the change in time as the period $dt$ approaches zero. We can make $n$ sufficiently large to use the tools of calculus to accurately approximate results, but in doing so we must remember that $\tau$ never approaches zero to the point where a beginning and ending time cannot be distinguished (i.e., $t_A \neq t_B$ for the same period no mater how small). Each time slice no matter how small has a beginning and an end time with some relevant meaning to our examination of a financial market. Therefore, any activity, including the $\tau$-related derivatives of price, will end up being \textit{averages} from $t_\epsilon$ to $t_{\epsilon+1}.$\footnote{To be rigorous would require special notation to denote these averages, such as $\left<\cdot\right>,$ however, such notation is unnecessarily cumbersome given this clarification of what we must consider an average and why.}

From this we see that $t=n\tau$
if all slices are equal in interval. If the intervals are not equal then
\begin{displaymath}
t=\sum_{\epsilon=A}^B \tau_\epsilon.
\end{displaymath}

Since we can arbitrarily define the period $t$ (e.g., one year, one month, one week, one day, one hour, one second, etc.) it is sufficient to set $t=1$ as an arbitrary temporal unit as long as we are clear on the definition of that unit in practice and consistent throughout our examination. 

Consequently, we can discuss the mechanics of a financial market for any slice using that slice as a unit of time. Although we get multiple units if we sum up the unit slices, we can always covert that sum itself into another unit slice as long as our terms are well-defined. Accordingly, as long as we are always comparing the same elapsed periods (e.g., one-week periods from 1 January 1980 to 31 December 2016, one-minute periods from 1 January 1980 to 31 December 2016, etc.), we can regard each period as a single unit for purposes of the elapsed time, or $t=1$.

\subsection{Displaced Price}

The question then becomes, ``What is the most basic displacement dynamic occurring over each time slice $\tau$?'' From an observation of assets traded on public exchanges it appears that the most basic displacement dynamic at each time slice is that the asset is either moving up or down in price to varying degrees.\footnote{It can also have no net movement. Since direction matters, price displacement and all functions of it are vector quantities even though we use scalar notation throughout with direction signified by positive or negative signs.} 

Let's call an ``up'' movement in price positive and a ``down'' movement negative. These are arbitrary but intuitive directional assignments. Again, as an asset moves in this way it travels not through traditional space but through the dimension of price. Let's call this price displacement $x$ defined as the change in price from the asset's starting price $x_A$ to its ending price $x_B$. 

Yet, it would be insufficient to define the price displacement $x$ merely as  $x_B-x_A$ as we did with time since doing so would leave us with a lack of homogeneity for financial markets. Generally, elapsed time and price displacements are homogeneous even though objective time and prices are not. There is arguably a relevant difference between the dates 14 March, 18 March, 25 November, and 29 November. Yet, the five-day period between 14 March and 18 March is chronologically identical to the five-day period between 25 November and 29 November. This also applies to years, months, weeks, hours, minutes, seconds, etc. Therefore, elapsed time is homogeneous even though specific times might be distinguishable. 

The same is true for changes in price, although here we must be more careful than with our examination of changes in time. There is a significant difference between \$100 and \$50. The two are distinguishable in relevant ways and are, therefore, inhomogeneous. Yet, there is no difference between the displacements $(x_{A_1}= \$90, x_{B_1}=\$100)$ and $(x_{A_2}= \$40, x_{B_2}=\$50)$. In both instances the price displacements are $x=x_B-x_A=\$10$. In all relevant ways the two displacements appear indistinguishable and, therefore, homogeneous. Yet, we are concerned with changes in price of a financial market for \textit{all} times. It appears changes in price alone end up being inhomogeneous over extended periods.  

For example, the S\&P 500 Index in the United States has data going back decades. If one were to take data from, say, 1980 through 2017, a \$10 change in price in 1980 would not be equivalent to a \$10 change in price in 2017 since the change from 1980 came from an initial price very close to \$100 while the change from 2017 came from an initial price very close to \$2,500. The extremely low (or for some markets high) values of early data can excessively skew the results so as to leave price displacements for more current periods inhomogeneous with respect to price displacements for earlier data even if the actual monetary displacement is identical.

Therefore, in order to maintain homogeneity we must measure price displacement in terms of the \textit{ratio} of the change in price from $t_A$ to $t_B$ to the objective price at $t_A$. For this reason, we will define the change in price $x$ as $(x_B-x_A)/x_A$, or for any time slice
\begin{displaymath}
x(\tau_\epsilon) :=\frac{x_{\epsilon+1}(\tau_\epsilon)-x_{\epsilon}(\tau_\epsilon)}{x_{\epsilon}(\tau_\epsilon)}=\left(\frac{x_{\epsilon+1}(\tau_\epsilon)}{x_\epsilon(\tau_\epsilon)}-1\right)
\end{displaymath}
instead of $x=x_B-x_A$ only.

Note that the closing price for $\tau_\epsilon$ is also the opening price for $\tau_{\epsilon+1}$. This is from the observation that financial markets leave off and pick up in the exact same spot during a period of exchange closure. One might argue that the price at the ``opening bell'' is the true opening price, but this is inconsistent with our conception of time as a series of slices. The closing price for $\tau_\epsilon$ maintains its value \textit{until} the exchange re-opens for trading. At the opening bell, any price is not the opening price, but the first price after $\tau_{\epsilon+1}$ has begun. 

For this reason, the price displacement in terms of reported price is always $x(\tau_\epsilon) =x_{\epsilon+1}(\tau_\epsilon)-x_{\epsilon}(\tau_\epsilon)$, with the price displacement ratio computed as already discussed. Hereafter, any discussion of the ``price displacement'' will be synonymous with ``price displacement ratio'' for the sake of concision unless otherwise specified. 

\subsection{Homogeneous Coordinates}

If we imagine a coordinate system with $t$ as the abscissa and $x$ as the ordinate we get a standard-looking stock chart with time running from left to right and the price running up and down. A consequence of this coordinate representation is that price displacements can be thought of as functions of elapsed time, or $x(t)$.

Therefore, the \textit{changes} in price and time as discussed in \S\S~2A and 2B are invariant under translation, even though prices and time themselves are not. If the changes in price and time are homogeneous then they are also isotropic, meaning they are invariant under rotation.

For these reasons, we can always make both $t_B$ and $x_B$ equivalent to $t$ and $x$, respectively, if we regard $t_A$ and $x_A$ as zero in relation to $t_B$ and $x_B$, respectively, for any time slice. Therefore, in this coordinate system, we can assume four price-time principles: (1) translational invariance, (2) rotational invariance, (3) time translational invariance, and (4) Galilean invariance (the rules apply equally to all assets even if each is in its own coordinate system).\footnote{It is well known that Newton's Second Law is invariant under a Galilean transformation so we need not prove this again here. By the end of this paper it should be clear that the price displacement of any asset is subject to the same rules regardless of its individual coordinate system with the inertial coefficient being the main determinant of price displacement magnitude given equal net trading interactions.} These principles are important because they are the four invariances Wigner claims necessary for the \textit{discovery} of mathematical rules of a system \cite{wigner}. Without these invariances, mathematical rules might exist for a system, but we would be hard pressed to root them out. The perceived lack of these invariances in economic systems is a common criticism for applying physical laws, even if by analogy, to financial markets \cite{mccauley}.

As a consequence, we can represent the end state of any elapsed period with the coordinates $(t,x)$ and that should be sufficient to describe the remaining dynamics occurring over the time interval in terms of time and price.

The only difference in the coordinates is that $t$ will always be positive since time only moves in one direction for our purposes here, while $x$ might be positive, negative, or zero since prices move up, down, or experience no change. 

In this way, our coordinate system becomes one of net changes from state $A$ to state $B$ since we can always set $\epsilon$ to $A$ and $\epsilon+1$ to $B$ for any slice $\tau$ regardless of the actual elapsed time in conventional units. Because of this, we will refer to our coordinate system of changes in time and price that sets $t_A=x_A=0$ and $t=1$ as a ``homogeneous coordinate system,'' although the reader should be aware that others might define this term differently in the literature \cite{casse}.

\section{External Forces} \label{Postulate II.}

At this point we should be able to imagine the price displacement of an asset as the net effect of the interactions of buyers and sellers. In fact, based on our postulates, these interactions are the only source of a net ``force'' that moves the asset through the price dimension over a time slice $\tau.$ 

Are these interactions sometimes the result of traders reacting to external events such as a significant terrorist attack or news of a major corporate scandal? Of course, but neither the attack nor the scandal directly affect the mechanics of this system we call a financial market. It might be the individual reactions of traders to these events that in the aggregate affect the direction of the price displacement over some elapsed period, but it is important to remember that financial markets have only one ``prime mover'' when we examine the system \textit{qua} system. That mover is the net force generated by the interactions of traders. There is no other source of direct force in this theory.\footnote{This theory does not disregard the uniqueness of each individual agent's behavior. The difference is in focus. Instead of focusing on the individual \textit{agent}, this theory focuses on the \textit{interaction} of individual agents.}

We use the term ``force'' here only by analogy; however, we will assume the net force to have the same properties and produce the same effects on an asset in the dimension of price as a physical force has on an object in the dimension of space. Recall that we assume financial markets comply with certain principles of physics, even if we agree that financial markets are not subject to all physical laws (e.g., gravity).

We know from experience that buyers and sellers interact with an asset and each other for each time slice $\tau$ that an exchange is open for trading.\footnote{Some ``penny'' stocks might go for long periods without buyer or seller activity. The markets we examine here are those with high trade volumes for all relevant times. The only reason for there to not be a trade in these markets would be that the exchange is closed so the interactions cannot occur.} Let's call a buyer's activity $\alpha$ since the activity is physically manifested in what's called an ``ask'' and a seller's activity $\beta$ since the activity is physically manifested in what's called a ``bid.'' Because we've decided to make the up direction positive and the down direction negative, we can notate a buyer's activity as $+\alpha$ and a seller's activity as $-\beta$. 

We assume that the price displacement for some period $\tau$ is proportional to the sum of the buyers' and sellers' activities, or 
\begin{displaymath}
x \propto \alpha + (-\beta)=\alpha-\beta,
\end{displaymath}
since these are the only activities that affect the asset's movement through the price dimension. Consequently, if $\alpha=\beta$ then the actions cancel, producing a price displacement $x=0$ for that period. However, if $\alpha > \beta$ then $x>0$, and if $\alpha<\beta$ then $x<0$. 

More specifically, we can think of the interactive dynamic $\alpha -\beta$ as generating some net force that is solely responsible for the price displacement we witness for the period. In other words, the net force is proportional to the change in price, or $F \propto x$. 

Here, $F$ is the net force resulting from the superposition of $F_\alpha$ and $F_\beta$ where $F_\alpha$ is positively directed and $F_\beta$ is negatively directed, or 
\begin{displaymath}
F=F_\alpha + (-F_\beta)=F_\alpha -F_\beta.
\end{displaymath}
The system as we have defined it appears in all respects to be linear. Therefore, it is appropriate to assume that the net response of the system to the interactions of traders is the sum of the responses that the buyers' and sellers' activities would have caused separately.

We can also assume that something ``carries'' these forces to produce the observed price displacement of an asset. For now, we will refer to this net ``carrying thing'' as $\psi$, where $\psi$ is the result of the superposition of the ``carrying thing'' of the buyers' force and that of the sellers'.

Let's return to our proportional assumption that $F \propto x$. Since both $F$ and $x$ are time functions, we know per Newton's second law that $F=m \ddot{x}$, where $m$ can be thought of as an inertial coefficient unique to each asset and $\ddot{x}$ is the second time derivative of the price displacement. This leaves us with $m \ddot{x} \propto x$.

If $m\ddot{x}$ displaces the asset in the price dimension over, let's say, the period $\tau_1$ it means the equivalent of $t_A$ is $t_1$ and the asset moves from price $x_1$ to price $x_2$ at $t_2.$ For the sake of argument let's assume the displacement over this slice is in the positive direction. Yet, at precisely $t_2$ the price $x_2$ that was the close price for the previous slice $\tau_1$ becomes the opening price for the next slice $\tau_2$. Our homogeneous coordinate system is such that $t_A$ and $x_A$ can always be considered zero since we care only about the net displacement over the period. For this reason, when $x_2$ goes from being the closing price for $\tau_1$ to being the opening price for $\tau_2$ the asset must reclaim the value zero since we regard the starting price for any slice as zero. 

By doing this, one of two things happen (both of which are equivalent): either the asset moves from the price $x_2=x_2$ at the end of $\tau_1$ to the price $x_2=0$ at the start of $\tau_2$, or the coordinate system moves to set $x_2=0$ at the start of $\tau_2$. Either way, the price displacement for the first slice, $x(\tau_1)$, must be traversed again in full, but in the opposite direction, so that $x_2=0$ for the start of $\tau_2$. This means if $m\ddot{x}$ is the net force that displaces the asset in the positive direction over $\tau_1$, there must be a negatively-directed force of equal magnitude returning price $x_2$ to zero for the start of $\tau_2.$ We'll express this negatively-directed force as $kx$.

This turns our proportional statement $m \ddot{x} \propto x$ into
\begin{equation} \label{eqn:oscillator}
m \ddot{x} = -kx.
\end{equation}
Since $m\ddot{x}$ is a positively-directed force and $kx$ is a negatively-directed force of equal magnitude, we must add a negative sign to $kx$ so that the two sides of the equation balance.

This is a very familiar equation. It describes the motion of a simple harmonic oscillator.\footnote{More generally, the system can be expressed as a Van der Pol equation $\ddot{x}-\gamma(1-x^2)\dot{x}+x=0$ with the constant $\gamma$ at or very near zero keeping the limit cycle close to circular since, as we shall see in \S~4, the theory relies on either the total or very near total conservation of energy with any net energy introduced into the system wholly converted to motion in the price dimension.} This is consistent with observations of a financial market from the perspective of the homogeneous coordinate system we have created. At each time slice $\tau$ the basic dynamic is that the asset moves through the price dimension in one direction and then must return (or the coordinate system reset) to the equilibrium point for the next slice thereby covering the same distance in price.

So far our coordinate system has the dimensions of time and price. Yet, there are well-known solutions to Eq. \eqref{eqn:oscillator} in the complex plane. If a solution to an equation of motion derived from our observations of how a financial market moves in price resides in the complex plane, we must assume that the complex plane is a necessary component of the lens through which we observe this real phenomenon even if it ends up being an intermediate mathematical contrivance. 

To observe this solution we must add one more dimension---an imaginary dimension---to our coordinate system. This third dimension is represented by the imaginary-axis $\imaginary$, thereby making our price dimension equivalent to the real-axis $\real$. Let's rotate the coordinate system so that the positive time-axis is heading into the page, the positive $\real$-axis is pointing to the top of the page, and the positive $\imaginary$-axis is pointing to the right of the page. We're now observing only the two dimensions of the complex plane. This configuration is slightly unconventional, but it maintains the visual of positive price changes going ``up'' and negative changes going ``down.''

We find a complex solution for Eq. \eqref{eqn:oscillator} in Quadrant I of the complex plane (upper right) in the following form:
\begin{equation} \label{eqn:complex number}
\psi_B=\mathcal{R}\left(\cos \sqrt{\frac{k}{m}}t + i\sin \sqrt{\frac{k}{m}}t \right),
\end{equation}
where $\mathcal{R}$ is the modulus of $\psi_B$.

The principal value of the argument of the complex number $\psi$ at $t_B$, or $\arg (\psi_B)$, is $\phi_B$. Let's call $\phi_\epsilon$ the phase of system, where the phase is defined geometrically in the complex plane as the angle from the positive $\real$-axis to a vector rotated toward the $\imaginary$-axis. In our coordinate system, this means the phase is positively directed if it moves clockwise from the $+\real$-axis towards the $+\imaginary$ axis.\footnote{Most configurations in the literature have the positively-directed phase moving counterclockwise. This is simply a result of how one chooses to set up the coordinate system, so the variation is irrelevant for our examination.}

Reviewing the complex solution for Eq. \eqref{eqn:oscillator} we hold that the phase $\phi_\epsilon$ is equivalent to $t\sqrt{k/m}$. Let's denote the \textit{change} in the phase from $t_A$ to $t_B$ as
\begin{displaymath}
\phi := \Delta \phi_\epsilon=\phi_B-\phi_A.
\end{displaymath}

Defined this way, at $t_A=0$ where $x_A=0$, $\phi_A$ must always equal $\frac{\pi}{2}$. This means
\begin{displaymath}
\phi=\phi_B-\frac{\pi}{2}.
\end{displaymath}
Thus, we see that the principle value of the phase displacement $\phi$ is always the complement of the principle value of the phase at $\phi_B$. In Quadrant I, the phase displacement $\phi$ is always negatively directed (rotates counterclockwise) with the only exception being $\phi_B=\phi_A=0$. 

If we span all quadrants for each $\tau$ we see that the range of $\phi_B$ is $\{ -\pi,\pi \}$ from its zero point along the $\real$-axis, consistent with Eq. \eqref{eqn:complex number}, and the range of $\phi$ is $\{ -\frac{\pi}{2},\frac{\pi}{2} \}$ from its zero point along the $\imaginary$-axis. Again, $\phi_A$ is always $\frac{\pi}{2}$ along the $\imaginary$-axis.

The phase displacement can be considered a \textit{functional} of the price displacement, or $\phi[x(t)]$. This simply means that the phase displacement is a number, the value of which is entirely dependent on the price displacement, which is a function with elapsed time as its parameter. As we shall see later in our discussion of the action (which is also a functional of the price displacement), the price derivative for which the functional derivative is always zero is an extremum.

From this we can express the complex number in Eq. \eqref{eqn:complex number} in terms of the phase displacement $\phi$ with the following:
\begin{equation} \label{eqn:complex sol}
\psi_B=\mathcal{R}(\sin \phi + i\cos \phi).
\end{equation}
The real solution for Eq. \eqref{eqn:complex sol} (the price displacement we actually observe) becomes 
\begin{equation} \label{eqn:real sol}
x=\mathcal{R}\sin \phi.
\end{equation}

The real solution for the price displacement in Eq. \eqref{eqn:real sol} is typically a result of our solving Eq. \eqref{eqn:oscillator} and ignoring the imaginary parts of the complex solution $\psi$ in Eq. \eqref{eqn:complex sol}. Taken as a physical description, this approach implies that we only observe one part of a much richer physical reality, the imaginary part of which somehow still lingers in existence but beyond our ability to observe it. Such an approach certainly stirs the imagination to conceive of a ``hidden dimension'' that exists beyond our senses yet is still responsible for the phenomena we experience in real-world financial markets. 

As fascinating as the implications of this approach might be, there is an explanation besides that of a lingering hidden dimension that is just as mathematically legitimate; one that regards the imaginary elements of the complex plane as a mere mathematical tool to help us understand the reality we see, but not as a physical reality itself. Let's examine an approach that yields the same result as Eq. \eqref{eqn:real sol} for an asset's price displacement without admitting a lingering hidden dimension in our physical reality.

So far we have only discussed $\psi$ as the solution for Eq. \eqref{eqn:oscillator}. Mathematically, there also exists a solution for Eq. \eqref{eqn:oscillator} in Quadrant II (upper left) that is the mirror image of $\psi$ reflected about the $\real$-axis:
\begin{displaymath}
\psi^*=\mathcal{R}(\sin \phi - i\cos \phi).
\end{displaymath}
This is the complex conjugate of $\psi$.

Since $\psi^*$ is a mirror reflection of $\psi$, we can assume both have the same phase displacement magnitude of $\phi$. Therefore, we can consider the price displacement $x$ to be the one-dimensional median between $\psi$ and $\psi^*$ in the complex plane. Because the distribution of planar area is symmetric between $\psi$ and $\psi^*$, we can express this median as
\begin{equation} \label{eqn:psi average}
x=\frac{1}{2}(\psi + \psi^*).
\end{equation}
We can say the same for $x$ in the negative direction using $-\psi$ and $-\psi^*$. 

Given our definitions of $\psi$ and $\psi^*$ we see that
\begin{displaymath}
\psi + \psi^* = 2\mathcal{R}\sin \phi.
\end{displaymath}
Note that the imaginary terms $i \cos \phi$ in $\psi$ and $-i \cos \phi$ in $\psi^*$ cancel when added. Substituting this sum into Eq. \eqref{eqn:psi average} we recover Eq. \eqref{eqn:real sol}.

By thinking of the price displacement as the one-dimensional median of the two-dimensional region between the wave function and its complex conjugate, we arrive at the same solution using a mathematical contrivance that cancels the imaginary terms instead of admitting a lingering dimension that physically exists but which we simply ignore. It is not merely ``more complete'' to regard the price displacement this way, it's necessary. Simply ignoring the imaginary parts deprives us of a fuller understanding of how the price displacement, the probabilities of these displacements, and the action relate to one another.

Because we are dealing with a sinusoid we can define the average frequency $\nu$ in cycles per unit of elapsed time and the average phase displacement per unit of elapsed time from states $A$ to $B$ as $\omega$, where
\begin{equation} \label{frequencies}
\omega=\frac{\phi}{t}=2\pi\nu.
\end{equation}
We will return to these terms later in our discussion of probabilities and our approximation for the extreme price displacements $\pm \mathcal{R}.$

We now have the elements of our three-dimensional homogeneous coordinate system: elapsed time, displaced price, and displaced phase, or $(t,x,\phi).$ From these three coordinates we can determine all relevant dynamics of the financial market that occur over the period $t$. Note that the price displacement is a function of elapsed time and the phase displacement is a functional of the price displacement. This gives us the homogeneous coordinates $(t:x(t):\phi[x(t)]),$ with each coordinate having some proportionality to a unit of elapsed time, or
\begin{displaymath}
\left(\frac{t}{t}:\frac{x}{t}:\frac{\phi}{t}\right). 
\end{displaymath}
Since we can regard $t=1$ with our unit assumption, we can reduce this to a two-dimensional coordinate system with no loss in generality. This gives us the simplified homogeneous coordinates $(x:\phi)$ that become equivalent to the ``speed'' of the market's mechanics over each unit of elapsed time, or $(\dot{x}:\omega)$.

Set up this way, $\psi=\psi_A+\psi_B$ is the complex superposition of things that carry the forces of buyers and sellers with a resulting amplitude of $\pm \mathcal{R}$. This is consistent with our assumption that some ``carrying thing'' is responsible for transporting the net force generated by the trading interactions of buyers and sellers that ultimately effects a price change for an asset. As with many things in the physical world, the thing that carries the force in our theory is akin to a wave, or at least its analog.\footnote{Although it is more accurate to refer to these as \textit{wave-like functions}, for the sake of brevity we will call them \textit{wave functions} throughout.}

Throughout this paper we are ultimately interested in the price displacement $x$, which we know is equivalent to $x_B$ given that $x_A$ can always be considered zero. Yet, $\psi_A$ cannot be considered zero since $\psi_A=i \mathcal{R}$, so $\psi \neq \psi_B$. Still, we will use $\psi$ to denote $\psi_B$ for the remainder of this paper with the clear understanding that when we use $\psi$ hereafter we are really talking about the complex number $\psi$ at state $B$ and not the change in the complex numbers from states $A$ to $B$.

Looking at our solution for the price displacement in Eq. \eqref{eqn:real sol} we can make two related observations. The first is that there exists a unique phase displacement measure $\phi$ for each possible price displacement over period $t$. The second is that the net movement in price over period $t$ is some maximum absolute measure multiplied by a function that oscillates between $-1$ and $+1$. In other words,  $x_{\min} =-\mathcal{R}$ and $x_{\max} =+\mathcal{R}$. 

This has significant implications for our theory of market mechanics as it suggests there is some net price displacement measure beyond which an asset cannot go for a defined period. The absolute price displacement can be greater than $\mathcal{R}$ for a period less than $t$ (e.g., $t_{n-j}- t_A$ where $j$ is a positive, non-zero integer); however, it cannot be greater than $\mathcal{R}$ for $t_B - t_A$.  This means an asset's net price movement is \textit{constrained} for a specified period, not by some external regulation, but by the properties of the asset itself. We will cover this more fully in a later section.


\section{The Action Principle} \label{Postulate III}

Our next endeavor is to examine whether the motion of an asset through price is consistent with the principle of stationary action, meaning we will investigate whether the action of the market system is stationary under small perturbations along its path from $x_A$ at $t_A$ to $x_B$ at $t_B$, at least to a first order approximation \cite{hamilton1}\cite{hamilton2}.

We would be remiss if we did not acknowledge that the discussion in this section of a financial market as a system with ``energies'' (even a discussion by analogy) understandably appears as a Samuelsonian nightmare for some readers \cite{samuelson}. Please keep in mind that, per our first postulate, we use these concepts analogously but hold that the analogous concepts might help us better understand the dynamics of the market mechanics we observe.

We do not make this examination for its own sake. We will use the calculated value of the market's action to derive a method for approximating the constraints of the system's price displacement for any time slice that are a consequence of its oscillating mechanics.

\subsection{The Action and the Lagrangian Generally}

The Lagrangian is a way of describing a financial market as a function of the conditions at $t_\epsilon$ (or any initial time for a slice $\tau$) relating to the price at the beginning of the time slice ($x_\epsilon$) and the first derivative of that price ($\dot{x}_\epsilon$). It contains the complete information of both the system and the effects of forces acting upon the system.\footnote{Dirac would most likely consider the Lagrangian not as a function of the asset's initial coordinates and its first derivative, but instead as a function of the asset's price at time $t_\epsilon$ and its price at time $t_{\epsilon+1}$, i.e., as endpoints \cite{dirac}.} 

Alternatively, the Lagrangian is often defined as the difference between the kinetic term $K$ and the potential term $V$ of a system, each measured at $t_\epsilon$, since $K$ expresses a property in terms of $\dot{x}_\epsilon$ and $V$ in terms of $x_\epsilon$. Therefore, $\mathcal{L}=K_\epsilon-V_\epsilon$.\footnote{Note that the kinetic term is simply the integral of the left-hand side of Eq. \eqref{eqn:oscillator} and the potential term is the negative integral of the right-hand side, both with respect to price displacement.} It is well known that in any physical system the observable path of an object (the trajectory the object actually takes through configuration space over time) minimizes this differential $\mathcal{L}$ over time.\footnote{This is shown by the Euler-Lagrange equation.} This is where the term ``principle of least action'' comes from.

Still speaking analogously, recall from the third postulate that this theory posits that an asset contains none of its own ``force'' and, therefore, none of its own ``energy.'' Any displacement of the asset in price is a result of work done on the asset, which is equal to the net external force over the amount of any displacement. The theory further holds that there is only one aggregate source that generates this net external force and introduces it to the asset: the net effect of buyers' and sellers' trading interactions.

From the previous sections we see that at each time slice $\tau$ the asset moves linearly in price. As applied here, the principle of stationary action holds that for each elapsed time $t$ the ``path'' taken by an asset (i.e., the curve traced out in the relative configuration space of the price dimension) between times $t_A$ and $t_B$ is the one for which the action does not change (i.e., is stationary) under small changes in the relative configuration of the asset related to the relative price dimension.

Since the price dimension is our vertical coordinate axis, we can go further and state that such a path is a straight line in either the up or down direction, although under rotation we can generalize it as the positive or negative direction. This is consistent with the oscillating dynamic expressed earlier.

It is important to note this is only consistent with observed market mechanics when $t=\tau$; that is, the asset moves strictly up or down in price only at local time slices. What is ``local?'' Well, that depends on what measure we wish to give $n$ and our definition of a temporal interval unit. 

Thought of another way, the action---again, the time integral of the Lagrangian---over any region of our price-time coordinate system must be stationary for any small changes in the coordinates in that region. If we keep dividing the regions until we get to a collection of time slices, we observe this stationary action as the binary ``up-down'' oscillations of typical market mechanics at each time slice $\tau$, which again is the elapsed time from $t_\epsilon$ to $t_{\epsilon+1}$. Therefore, for each time slice in our price-time coordinate system there exists a Lagrangian that is a function of its coordinates and their first derivatives with respect to time.

Does this mean we can't determine the action between $t_A$ and $t_B$ when the elapsed time is not local? No, in fact, quite the opposite. It simply means we must first examine the specific dynamics at each slice $\tau$ between $t_A$ and $t_B$ and then sum over all periods $\tau$ to find measurable results that match observed data. In other words, the action $\mathcal{S}$ from $t_A$ to $t_B$ is equal to the following series of time slices:
\begin{equation} \label{adding Lagrangian}
\int_{t_A}^{t_B} \mathcal{L}dt=\int_{t_0}^{t_1} \mathcal{L}dt+\int_{t_1}^{t_2} \mathcal{L}dt+\cdots+\int_{t_{n-1}}^{t_n} \mathcal{L}dt,
\end{equation}
where $\tau_\epsilon=t_{\epsilon+1}-t_\epsilon$ just as on a larger scale we find $t=t_B-t_A.$\footnote{Of course, we can always set $t$ to $\tau_\epsilon$, $t_B$ to $t_{\epsilon+1}$, and $t_A$ to $t_\epsilon$ if we choose.}

The Lagrangian over each time slice $\tau_\epsilon$ in our price-time coordinate system makes some contribution to the total price displacement $x$ measured between $t_A$ and $t_B$. This is similar to Dirac's approach \cite{dirac} and, later, Feynman's definition \cite{feynman}. Observe, however, that the price displacement is not only a result of the phase displacement of $\psi$. It is also a result of the phase displacement belonging to the complex conjugate of $\psi^*$, which we'll denote here as $\phi^*$ to avoid confusion.\footnote{Technically, the complex wave function $\psi$ has the conjugate $\psi^*$ and both have identical phase displacements $\phi$, although they are oppositely directed; however, it's easier to talk about the phase displacements attributed to each by denoting them, for the moment, as $\phi$ and $\phi^*$.} In fact, both $\phi$ and $\phi^*$ contribute in equal measure to $x$ since, as we showed earlier, a complex number and its complex conjugate share the same $\real$-axis value $x_B$, which in our homogeneous coordinate system is equivalent to the price displacement $x$. So if the Lagrangian for each time slice $\tau$ makes some contribution to the total price displacement $x$ measured between $t_A$ and $t_B$ so does \textit{twice} the phase displacement for that time slice, which we'll denote as a functional of the price displacement for that interval $\tau$, or $2\phi[x(\tau)]$. 

Since the local elements of Eq. \eqref{adding Lagrangian} sum to the total time integral of the Lagrangian between $t_A$ and $t_B$, and therefore the total action, we can state that each element $\int_{t_\epsilon}^{t_{\epsilon+1}} \mathcal{L}~dt \equiv d\mathcal{S}$ (if $n$ is sufficiently large) and that each element is minimized under the principle of stationary action. 

\subsection{Measuring the Action}

We begin measuring the action by examining the potential and kinetic terms of the system. Any positively-directed force can be expressed as the negative spatial derivative of the system's potential. Thus, by analogy, we can express the negatively-directed force $kx$ from Eq. \eqref{eqn:oscillator} as the positive price derivative of the potential (or its analog) of the system we call a financial market. This potential $V$ then becomes
\begin{equation} \label{V}
V=\frac{1}{2}kx^2.
\end{equation}
Note that the potential term of the asset is simply the price integral of the right-hand side of Eq. \eqref{eqn:oscillator}. The kinetic term should then be the price integral of the left-hand side, which becomes
\begin{equation} \label{K}
K=\frac{1}{2}m\dot{x}^2.
\end{equation}

At this point we should already begin to suspect an equivalence between the potential and kinetic terms in financial markets since both are results of price integrals of the balance of the net interaction of buyers and sellers expressed in Eq. \eqref{eqn:oscillator}. Still, we will dive more deeply into this over the coming pages in order to be sure.

The next question becomes, ``What is the measure of this action, and by extension, the measure of the phase displacement?'' We attempt to answer this by again examining the dynamics at each time slice.

At each slice $\tau$ the net external force causes a tiny amount of work to be done with respect to a tiny displacement in price, especially when $n$ is very large, or $F(\tau_\epsilon)=\frac{dW}{dx(\tau_\epsilon)}$. To figure out the total amount of work done on the asset over the period  we take the price integral of the net external force, or $W=\frac{1}{2}Fx$.

The work $W$ is equivalent to the total potential necessary to be introduced into the asset from the net interactions of buyers and sellers in order to make the asset move in price. Since the asset neither has nor stores any of its own ``energy,'' any potential used to move the asset through the price dimension must come from the net potential introduced into the asset by these external interactions. In other words, the kinetic term $K$ resulting from any price displacement must come from an external potential $V$. This means the final kinetic term must equal the initial potential as measured over any elapsed period. 

Consequently, we can hold that the total work performed on the asset equals the total potential initially introduced into the asset. This is also equal to the final kinetic term from the movement of the asset through the price dimension. In other words, $W=V_\epsilon=K_{\epsilon+1}$. For this to be true, $V_{\epsilon+1}$ and $K_\epsilon$ must equal zero, with $V$ and $K$ ``trading off'' measures over each elapsed period but the sum of $V$ and $K$ always equaling $W$ for that time slice. 

So the answer to the question, ``What is the potential of the system?'' depends entirely on when we ask the question over the elapsed time slice $\tau.$ For this reason, it becomes impossible to express the potential and kinetic terms in any way except as averages over $\tau.$ The only exceptions are expressions of the potential and kinetic terms at the endpoints $t_\epsilon$ and $t_{\epsilon+1}$ of the generic slice $\tau_\epsilon.$ Since $K_\epsilon=0$, the Lagrangian becomes 
\begin{displaymath}
\mathcal{L}=K_\epsilon-V_\epsilon=0-V_\epsilon=-V_\epsilon.
\end{displaymath}

We've made a bit of progress. We now know that the Lagrangian is the negative initial potential generated by the interactions of buyers and sellers for any time slice $\tau_\epsilon$. But what is the measure of this potential? We know it must be the same as the average measure of work, which is equal to $\frac{1}{2}Fx$. We also know from Newton's second law that $F=m\ddot{x}=\frac{mx}{t^2}$ since $\ddot{x}=x/t^2$ when taken as an average. Therefore,
\begin{displaymath}
\frac{1}{2}Fx=\frac{mx}{2t^2}x=\frac{mx^2}{2t^2}.
\end{displaymath}

But this is the same as the measure of the average kinetic term of a system since the average $\dot{x}$ is always $x/t$. It seems, therefore, that $K_{\max}=V_{\max}$. Since $K_{\max}$ implies that $V_{\min}=0$, $V_{\max}$ implies that $K_{\min}=0$, and $V_\epsilon \equiv V_{\max}$, we can conclude that 
\begin{displaymath}
V_\epsilon=\frac{k x^2}{2}=\frac{mx^2}{2t^2}
\end{displaymath}
as an average measure for $t=\tau_\epsilon$.\footnote{There are some who might argue that if we are expressing the average kinetic term where $\tau$ is small we must do so as 
\begin{displaymath}
\frac{m}{2}\left( \frac{x_{\epsilon+1}-x_\epsilon}{t_{\epsilon+1}-t_\epsilon} \cdot \frac{x_{\epsilon}-x_{\epsilon-1}}{t_{\epsilon}-t_{\epsilon-1}} \right)
\end{displaymath}
not as $\frac{m}{2}\cdot \left(\frac{x}{t}\right)^2$ if the mean value of the square of the price displacement is  $dt$ instead of $dt^2.$ While this exception is noted, the notation used in this paper seems adequate for purposes of examining the price displacement of financial markets.}
Consequently, we can express the Lagrangian of an asset as 
\begin{displaymath}
\mathcal{L}=-V_\epsilon=-\frac{mx^2}{2t^2}.
\end{displaymath}

Because the action is defined as the time integral of the Lagrangian, we can express the action as
\begin{equation} \label{eqn:S}
\mathcal{S}=\int_{t_A}^{t_B} -\frac{mx^2}{2t^2}~dt=\frac{mx^2}{2t}.
\end{equation}

We can also find the measure of the action by calculating it from the theoretical Lagrangian. Expressed this way, the action becomes
\begin{displaymath}
\mathcal{S}=\frac{m}{2} \left(\int_{t_A}^{t_B} \dot{x}^2-\frac{k}{m} x^2 ~dt \right).
\end{displaymath}
Integrating by parts gives us the following related to the kinetic term:
\begin{displaymath}
\int_{t_A}^{t_B} \dot{x}^2~dt=\int_{t_A}^{t_B} \dot{x}\dot{x}~dt=\left| x\dot{x}\right|_{t_A}^{t_B}.
\end{displaymath}
Since $\ddot{x}=-\frac{kx}{m}$, we get
\begin{displaymath}
\int_{t_A}^{t_B} \dot{x}^2~dt=\left| x\dot{x}\right|_{t_A}^{t_B}+\frac{k}{m}\int_{t_A}^{t_B} xx~dt.
\end{displaymath}
The action then becomes
\begin{displaymath}
\mathcal{S}=\frac{m}{2}\left[\left| x\dot{x}\right|_{t_A}^{t_B}+\frac{k}{m}\int_{t_A}^{t_B} xx~dt-\frac{k}{m}\int_{t_A}^{t_B} xx~dt\right],
\end{displaymath}
which simplifies to 
\begin{displaymath}
\mathcal{S}=\frac{m}{2}\left| x\dot{x}\right|_{t_A}^{t_B}=\frac{m}{2}(x_B\dot{x}_B-x_A\dot{x}_A).
\end{displaymath}
Because we set $x_A=0$ in our homogeneous coordinate system, the $x_A$ terms disappear. Further, because we can only talk about the kinetic term over the period $\tau_\epsilon$ as an average, we get 
\begin{displaymath}
\mathcal{S}=\left(\frac{mx}{2}\right)\left(\frac{x}{t}\right)=\frac{mx^2}{2t},
\end{displaymath}
which is the same value we derived originally.

Recall the solution for the asset's price path for any elapsed period is $x(t)=1/2(\psi+\psi^*).$ We next want to know if the action along this path $x(t)$ is stationary, or more specifically, minimized. 

Assume $x(t)$ is the actual path the asset takes through the price dimension from $t_A$ to $t_B$, and $\xi(t)$ is some arbitrary path it can take between the two temporal endpoints. We'll introduce the term $\eta(t)$ as some variation from the actual path $x(t)$, where $\eta(t) :=\eta(t_A)=\eta(t_B)=0.$ We can then define the arbitrary path as $\xi(t) := x(t)+\eta(t).$

Since the Lagrangian is quadratic in $x$ and $\dot{x}$, the action for the arbitrary path $\xi(t)$ becomes
\begin{displaymath}
\mathcal{S}(\xi)=\mathcal{S}(x+\eta)=\mathcal{S}(x)+\mathcal{S}(\eta),
\end{displaymath}
which is valid for an oscillating system. Therefore, $\mathcal{S} > 0$ when $\eta(t_A)=\eta(t_B)=0.$

The action of the arbitrary system then becomes
\begin{displaymath}
\mathcal{S}(x)+\mathcal{S}(\eta)+\frac{m}{2}\left( \int_{t_A}^{t_B} \dot{x}\dot{\eta}~dt-\frac{k}{m} \int_{t_A}^{t_B}x\eta ~dt\right).
\end{displaymath}
If we rewrite the term $\dot{x}\dot{\eta}$ in terms of $\eta$, we find that the integrated terms disappear since $\eta=0.$ Therefore, $\mathcal{S}(\eta)=0$ meets the condition that $\mathcal{S} > 0$ when $\eta(t_A)=\eta(t_B)=0.$

We can also define the action as stationary if there is no change in $\mathcal{S}$ in the first approximation. In general, if we have a minimum quantity of a function then any perturbation away from that minimum in the first order will only have a deviation in the second order. At any other place besides the minimum a small perturbation will show up in the first order, but at the minimum any small perturbation will make no difference in the first approximation.

We can examine this by looking at the action as a series. The action $\mathcal{S}=\frac{mx^2}{2t}$ can be expressed as the infinite series
\begin{equation}
\sum_{n=0}^\infty \frac{1}{2}(t-1)^n \left[(-1)^n mx^2\right]
\end{equation}
for $|1-t|=0$, which is what we're looking for since we want the action to be stationary for any unit of elapsed time. Because we only need to calculate the value to the first approximation to judge if the action is stationary we can restrict this series from $n \in \{0,\infty\}$ to $n \in \{0,1\}.$ This gives us
\begin{displaymath}
\frac{1}{2}(t-1)^0 \left[(-1)^0 mx^2\right] + \frac{1}{2}(t-1)^1 \left[(-1)^1 mx^2\right],
\end{displaymath}
which becomes
\begin{displaymath}
\frac{1}{2}(1) \left[(1) mx^2\right] + \frac{1}{2}(t-1) \left[(-1) mx^2\right].
\end{displaymath}
This reduces to
\begin{equation}
\frac{mx^2}{2}(2-t).
\end{equation}
Note that at $t=1$ this equation is equivalent to Eq. \eqref{eqn:S}. Thus, we see that $\delta\mathcal{S}=0$; the action remains stationary to the first approximation for any temporal unit. Therefore, we can conclude the differential represented by the Lagrangian over $t$ is minimized.

From Eq. \eqref{eqn:S} we see that the action is parabolic with regards to the price displacement. Therefore, we can look for solutions for the price displacement that maintain the action as stationary. One way to do this is to take the partial derivative of the action with respect to the price displacement and find solutions when the partial derivative is zero. Our hope is that the action remains stationary for all possible values of the price displacement, meaning from zero to $|\mathcal{R}|$.

For any positive price displacement on the complex plane ranging from zero to $|\mathcal{R}|$ the absolute value of the phase displacement is in the range $[0,\frac{\pi}{2}].$ It is the same for any negative price displacement. Therefore, if solutions to $|\phi|$ include these two extremes when the partial derivative of the action with respect to the phase displacement is equal to zero, we can reasonably assume the action is stationary for all values of the price displacement $|x|$.

Recall from Eq. \eqref{eqn:real sol} that $x=\mathcal{R}\sin\phi.$ Thus, we can set the partial derivative of the action with respect to the phase displacement equal to zero with the expression
\begin{equation}
\frac{\partial}{\partial\phi} \left(\frac{m(\mathcal{R}\sin\phi)^2}{2t}\right)=0.
\end{equation}
This becomes
\begin{equation}
\frac{m\mathcal{R}^2\sin\phi \cos\phi}{t}=0,
\end{equation}
which generates two possible solutions: $g\pi$ and $g\pi+\frac{\pi}{2}$, where $g$ is an integer scalar. We're interested in solutions at $g=0$ since all other solutions are simply positive or negative integer multiples. This leaves us with the solutions
\begin{equation} \label{3pi/2}
|\phi|=0 \quad \mathrm{and} \quad |\phi|=\frac{\pi}{2},
\end{equation}
which are the extremes of the absolute value of each phase displacement we just mentioned. 

Accordingly, we can conclude that all possible paths for the price displacement $x$ will result in a stationary action to a first order approximation. Therefore, the mechanics of financial markets viewed through the lens of our homogeneous coordinate system conform to the principle of least action for all possible price paths.


\section{Deriving Probabilities} \label{Deriving Probabilities}

We now understand that there exists a unique complex number $\psi$ and its complex conjugate $\psi^*$ for every discrete price displacement $x$ in the inclusive range $x\in [-\mathcal{R},\mathcal{R}]$. We can express $\psi$ as an exponential function of $\phi$ in the form
\begin{equation}
\psi=i\exp[-i\phi].
\end{equation}
This becomes a probability (Gaussian) function by simply squaring the exponent 
\begin{equation}
i\exp[(-i\phi)^2]=i\exp[-\phi^2]
\end{equation}
and it becomes normalized by dividing the Gaussian by the sum of all possible values of the Gaussian
\begin{equation}
\Pr(\psi)=\frac{i\exp[-\phi^2]}{i \int_{-\infty}^{\infty}\exp[-\phi^2]~d\phi}=Q_\psi~e^{-\phi^2},
\end{equation}
where $Q_\psi$ is the corresponding normalization constant for $\psi$ in terms of the phase displacement with the value
\begin{equation}
Q_\psi=\frac{1}{\sqrt{\pi}}.
\end{equation}
We perform a similar operation to get $\Pr(\psi^*)$.

We also understand from our discussion of the Lagrangian in Eq. \eqref{adding Lagrangian} that for each unique price displacement $x$ there exist two unique phase displacements $\phi$; one in Quadrant I and another in Quadrant II of the complex plane for positive price displacements, and one in Quadrant III and another in Quadrant IV for negative price displacements. 

In other words, for any $+x$ there exist $2|\phi|$ and the same for any $-x.$ If we define the absolute range of each $|\phi|$ to be $[0,\frac{\pi}{2}]$ as we did at the end of \S~4 then for any absolute price displacement $|x|$ there exist $4|\phi|$ since $|x| \equiv \pm x$.

Similarly, we find that the probability of the unique price displacement $x$ occurring is a function of \textit{both} $\psi$ and $\psi^*$. Since $x$ results from the phase displacements associated with $\psi$ and the phase displacements associated with $\psi^*, $ we must add the $\phi$-terms for both sides to find the probability of a specific value $X$, or $\Pr(x=X)$. 

Feynman \cite{feynman} gives a much more elaborate method for doing this by defining multiple regions and then splitting them up, but we can also view this as a basic probability problem. By asking, ``What is the probability of $x$ having the specific value $X$?,'' we're essentially asking, ``What is the probability of finding both $\psi$ and $\psi^*$ each with specific phase displacement values $\phi=\Phi$ where $\Phi$ is the unique phase displacement for $X$?'' Formally, this becomes $\Pr(\psi \cap \psi^*).$ We know how to solve this since it is well known that
\begin{displaymath}
\Pr(\psi \cap \psi^*)=\Pr(\psi) \cdot \Pr(\psi^*)
\end{displaymath}
if $\psi$ and $\psi^*$ are independent events, which they are since they are unique and separate from each other. Because this is the same as the probability of a specific price displacement since the phase displacements are functions (more properly functionals) of the the price displacement, it follows that
\begin{displaymath}
\Pr\left(x\right)=\Pr(\psi) \cdot \Pr(\psi^*).
\end{displaymath}
Because $\Pr(\psi) \equiv \Pr(\psi^*)$, this gives us
\begin{equation} \label{Prob}
\Pr(x)=\Pr(\psi)^2.
\end{equation}

Note that we are simply adding the phase displacements together. Since the phase displacements are in exponential forms we add by multiplying the two exponential functions, which in this case is equivalent to squaring the original exponential function because $\Pr(\psi) \equiv \Pr(\psi^*)$. This is because the product of a series of exponential functions is equal to the exponential function of the sum of a series of exponents, or 
\begin{displaymath}
\prod_j \exp \left( a_j \right)=\exp \left( \sum_j a_j \right).
\end{displaymath}

Determining the probability of the price displacement having the value $X$ in terms of the action for each elapsed period then becomes
\begin{equation}
\left(\frac{i\exp[-\mathcal{S}]}{i \int_{-\infty}^{\infty}\exp[-\mathcal{S}]~dx}\right)^2=\left(Q_x~e^{-\mathcal{S}}\right)^2
\end{equation}
where $Q_x$ is the corresponding normalization constant for the price displacement in terms of the action with the value
\begin{equation}
Q_x=\sqrt{\frac{m}{2\pi t}}.
\end{equation}

This approach implies that the total probability measure of the absolute price displacement $|x|$ for any region of relative configuration space with boundary coordinates $(|x|,|\phi|)$ is the product of two identically split regions of three-dimensional relative configuration space with elapsed time $t=1$. This is because the probabilities as we've constructed them are exponential functions relating to the absolute phase displacement $|\phi|$. The entire probability measure is the combination of two evenly-split regions with respect to $|\psi|$ and $|\psi^*|$ along the $\real$-axis of our homogeneous coordinate system for each time slice $\tau.$

In short, we must square the traditional probability function for $|\psi|$ effectively doubling the absolute phase displacement since there are two possible ways in the complex plane we can get the exact same price displacements for $x$, namely, $\psi$ and $\psi^*$. The same is true for $-x$ with $-\psi$ and $-\psi^*$. This approach does not violate the principle of unitarity since probability measures remain between zero and unity.\footnote{One might argue that this approach assumes that price displacements in the positive and negative directions are symmetric, which in most cases they are not. However, if we only measure the probabilities of the absolute price displacements and the corresponding absolute phase displacements, this asymmetry should not matter for our purposes.}

Recall our expressions of the frequencies in Eq. \eqref{frequencies} and our use of an elapsed time unit. The probability of the phase displacement being at least some value $\Phi$ is equivalent to the probability of the phase displacement being at least $2\pi \nu(\Phi),$ where $\nu(\Phi)$ is the average frequency at the specific phase displacement value $\Phi.$ We can express this as
\begin{equation}
\Pr(|\phi| \geq \Phi) \equiv \Pr(|\phi| \geq 2\pi \nu(\Phi)).
\end{equation}
We can also express this equation using the Gauss complementary error function as
\begin{equation}
\mathrm{erfc}(\Phi) \equiv \mathrm{erfc}(2\pi \nu(\Phi)), 
\end{equation}
which implies
\begin{equation}
\Phi = 2\pi \nu(\Phi)
\end{equation}
or
\begin{equation} \label{f}
\nu(\Phi) = \frac{\Phi}{2\pi}.
\end{equation}

Let's next try to express the probability of the absolute price displacement $|x|$ being at least the value $|X|$ in terms of the action $\mathcal{S}.$ We know the value of the action from Eq. \eqref{eqn:S}. To find the normalized probability of $|x| \geq |X|$ in terms of the action we integrate the action as an exponent from $|X|$ to infinity. We then normalize the result by dividing by the integral of the same from zero to infinity. This gives us
\begin{equation}
\frac{\int_{|X|}^\infty e^{-\mathcal{S}}dx}{\int_{0}^\infty e^{-\mathcal{S}}dx}=\mathrm{erfc}\left( \sqrt{\mathcal{
S}}\right).
\end{equation}
But recall that this is only one-half of the mechanics since it represents the probability in terms of $|\psi|$ only. To complete the picture, we must add the $|\psi^*|$ half, thereby squaring the entire quotient. This results in
\begin{equation}
\Pr(|x| \geq |X|) = \mathrm{erfc}\left( \sqrt{\mathcal{
S}}\right)^2.
\end{equation}
This implies
\begin{equation} \label{erfc}
\Pr(|x| \geq |X|)=\mathrm{erfc}\left(|X|\sqrt{\frac{m}{2t}}\right)^2
\end{equation}
from \eqref{eqn:S}. 

Since there exist $4|\phi|$ for any $|x|$, we can state further that
\begin{equation}
\Pr(|x| \geq |X|)=\mathrm{erfc}\left(4\Phi\right)^2=\mathrm{erfc}\left(8\pi \nu(\Phi)\right)^2.
\end{equation}
Combining these equations gives us
\begin{equation}
\mathrm{erfc}\left(|X|\sqrt{\frac{m}{2t}}\right)^2=\mathrm{erfc}\left(8\pi \nu(\Phi)\right)^2.
\end{equation}
This implies the general expression
\begin{equation} \label{general}
|X|\sqrt{\frac{m}{2t}}=8\pi \nu(\Phi).
\end{equation}

If $|X|=|\mathcal{R}|$ then $|\Phi|=\frac{\pi}{2}.$ From Eq. \eqref{f} we see that
\begin{equation}
\nu\left( \frac{\pi}{2} \right)=\frac{1}{4}.
\end{equation}
Substituting this into Eq. \eqref{general} we get
\begin{equation}
|\mathcal{R}|\sqrt{\frac{m}{2t}}=8\pi \nu\left( \frac{\pi}{2} \right) = 8\pi \left( \frac{1}{4}\right)=2\pi.
\end{equation}
This gives us a final expression for the measure of the extreme price displacement for any financial market as
\begin{equation}
|\mathcal{R}|=\pi\sqrt{\frac{8t}{m}},
\end{equation}
or, since $t=1$,
\begin{equation}
\mathcal{R}^2 = \frac{8\pi^2}{m}.
\end{equation}

If we then substitute this value into the square of our original solution for the price displacement in Eq. \eqref{eqn:real sol} we find that
\begin{equation}
x^2=\frac{8\pi^2}{m} \sin^2\phi.
\end{equation}
Rearranged, we can express the price displacement in Eq. \eqref{eqn:real sol} in the form of the action
\begin{equation}
\mathcal{S}=\left( 2\pi\sin\phi \right)^2.
\end{equation}
This implies the phase displacement is both a functional of the price displacement over an elapsed period and the inertial coefficient unique to each asset, or $\phi[x(t),m]$. So, too, is the action such a functional, which we already surmised from Eq. \eqref{eqn:S}.

Regarded this way, the action of a financial market's one-dimensional (linear) price path becomes the square of an analogous notion of two-dimensional (circular) rotational speed in the complex plane. We'll denote this two-dimensional quantity as $u.$ Therefore, we conclude that
\begin{equation}
\mathcal{S}=u^2.
\end{equation}

We will next apply these theoretical expressions to historical data from various financial markets and see to what extent the historical and theoretical results are consistent.


\section{Testing the Theory}
Now that we've laid the theoretical groundwork, let's see if the theory is, at a minimum, consistent with certain historical asset prices. The following describes our simple way to test for conformity with historical data, although I am certain there are others with much more robust methods to test this. 

The first part is really the set up where we compare theoretical probability calculations for the absolute price displacement being at least its historical value for each period against the relative frequency of that minimum displacement occurring over an extended period. If the probabilities have an acceptable correlation value (say, $r^2>0.99)$ it suggests the theory might not be \textit{in}correct (which is quite different, of course, from suggesting the theory might be correct). This first part will also allow us to approximate an average inertial coefficient for each individual market that we will use for our primary test.

The second part of the test for conformity is to predict the extreme price displacements for individual financial markets from data collected well in advance of, sometimes decades before, an historical ``crash'' and see if the predicted extreme is violated by the crash many years later. If it is not violated, the results suggest two things: that (1) again, the theory might not be incorrect; and (2) each financial market has an inertial coefficient (or its analog) that is an internal property of that asset.

Limited by time and access to free market data extending back multiple decades, we test here only a few financial markets for the unit period $t=1$ trading week. We use representative stock indices from American, European, and Asian exchanges, a commodity futures contract, a foreign currency exchange pair, and a single publicly-traded company. Each of these has sufficient trading volumes and freely accessible historical data reaching back into at least the 1990s and some into the early 1970s.

\subsection{Probabilities and Inertial Coefficients}

The first step in this test is to find weekly historical price data for the financial markets we wish to examine. The following are the weekly data that are freely available for each market from the date listed to the present:\footnote{Source: www.investing.com.}

\begin{itemize}
\item S\&P 500 (U.S.): 15 June 1980
\item Dow Jones 30 (U.S.): 10 February 1985
\item NASDAQ 100 (U.S.): 15 June 1980 
\item DAX (Germany): 03 January 1988
\item Nikkei 225 (Japan): 15 January 1984
\item WIG20 (Poland): 24 April 1994
\item Gold Futures: 27 January 1980
\item USD/JPY: 10 January 1971
\item Johnson \& Johnson (JNJ): 23 March 1980

\end{itemize}

The second step is to calculate the change in price from the previous week's close to the current week's close ($x_B-x_A$). We can then calculate the absolute price displacement ratio using $|x|=|x_B/x_A-1|$ for each week.

The third step is to code the data with either a ``0'' or ``1.''  A code of 0 means the value of $|x|$ for a particular week $w$ is less than some arbitrary value $|X|.$ A code of 1 means the value of $|x|$ is greater than or equal to $|X|.$ We can then take the average of all the 0's and 1's for each $|X|$ and call it the relative frequency of the market meeting the condition $|x| \geq |X|.$ We denote this relative frequency function $\rho.$

The fourth step is to approximate the average value of the inertial coefficient $m$ for each asset, which we denote as $\hat{m}.$ We do this by employing Eq. \eqref{erfc}, performing some algebraic manipulation, and concluding that the value of $m$ for a specific week $w$ is
\begin{equation}
m_w=2t\left( \frac{\mathrm{erfc}^{-1}(\sqrt{\rho(|x| \geq |X|)}}{|x|_w} \right)^2
\end{equation}
recalling, of course, that $t=1$ trading week. The approximation of the average value of the inertial coefficient for a market then becomes the value of $\hat{m}$ that yields the highest coefficient of determination value $r^2$ for the dependent variable $\Pr(|x| \geq |X|)$ and the independent variable $\rho(|x| \geq |X|)$  where $|X|=|x|_w.$

The following are the $\hat{m}$ and $r^2$ values for each asset using this method sampled from the first 100 weeks of the data set with the dates listed previously (e.g., the S\&P 500 sample is from the beginning of the week starting 15 June 1980 to the end of the week starting 16 May 1982):

\begin{itemize}
\item S\&P 500: $\hat{m} = 977.73$; $r^2=0.9992$
\item Dow Jones 30: $\hat{m} = 982.21$; $r^2=0.9997$
\item NASDAQ 100: $\hat{m} = 683.00$; $r^2=0.9988$
\item DAX: $\hat{m} = 504.66$; $r^2=0.9998$
\item Nikkei 225: $\hat{m} = 474.14$; $r^2=0.9990$
\item WIG20: $\hat{m} = 355.92$; $r^2=0.9988$
\item Gold Futures: $\hat{m} = 820.35$; $r^2=0.9994$
\item USD/JPY: $\hat{m} = 2513.76$; $r^2=0.9977$
\item JNJ: $\hat{m} = 511.80$; $r^2=0.9995$

\end{itemize}

Although we calculate the values of $\hat{m}$ for each market, the fact that the coefficient of determination is so high for most samples suggests that our method of calculation is very close to accurate.

\subsection{Extreme Price Displacements}

The real test for conformity of the theory with historical data is to then use the value of $\hat{\mu}$ in order to predict the extreme values for $|x|$ for any one week. This test requires the postulates and assertions of this theory be accurate. If the results are consistent, it suggests this view of financial markets through the lens of our homogeneous coordinate system may be of some value in future examinations of market mechanics.\medskip

\noindent S\&P 500 (SPX) \begin{itemize}
\item Data from: 1980-1982
\item Crash Week: 05 October 2008
\item Predicted $|\mathcal{R}|$ for Crash Week: 312.37
\item Actual $|x|$ for Crash Week: 200.01
\item Interval from Pred. to Crash: $\approx$ 26 years
\end{itemize}

\medskip

\noindent Dow 30 (DJI) \begin{itemize}
\item Data from: 1985-1987
\item Crash Week: 05 October 2008
\item Predicted $|\mathcal{R}|$ for Crash Week: 2927.51
\item Actual $|x|$ for Crash Week: 1874.19
\item Interval from Pred. to Crash: $\approx$ 21 years
\end{itemize}

\medskip

\noindent NASDAQ 100 (NDX) \begin{itemize}
\item Data from: 1980-1982
\item Crash Week: 09 April 2000
\item Predicted $|\mathcal{R}|$ for Crash Week: 1511.81
\item Actual $|x|$ for Crash Week: 1125.16
\item Interval from Pred. to Crash: $\approx$ 18 years
\end{itemize}

\medskip

\noindent DAX (GDAXI) \begin{itemize}
\item Data from: 1988-1990
\item Crash Week: 05 October 2008
\item Predicted $|\mathcal{R}|$ for Crash Week: 2292.98
\item Actual $|x|$ for Crash Week: 1252.72
\item Interval from Pred. to Crash: $\approx$ 18 years
\end{itemize}

\medskip

\noindent Nikkei 225 (N225) \begin{itemize}
\item Data from: 1984-1986
\item Crash Week: 11 October 1987
\item Predicted $|\mathcal{R}|$ for Crash Week: 6732.38
\item Actual $|x|$ for Crash Week: 3068.00
\item Interval from Pred. to Crash: $\approx$ 1 year
\end{itemize}

\medskip

\noindent WIG20 (WIG20) \begin{itemize}
\item Data from: 1994-1996
\item Crash Week: 05 October 2008
\item Predicted $|\mathcal{R}|$ for Crash Week: 1107.86
\item Actual $|x|$ for Crash Week: 360.54
\item Interval from Pred. to Crash: $\approx$ 12 years
\end{itemize}

\medskip

\noindent Gold Futures (GC) \begin{itemize}
\item Data from: 1980-1982
\item Crash Week: 18 September 2011
\item Predicted $|\mathcal{R}|$ for Crash Week: 562.18
\item Actual $|x|$ for Crash Week: 174.60
\item Interval from Pred. to Crash: $\approx$ 29 years
\end{itemize}

\medskip

\noindent U.S. Dollar Japanese Yen (USD/JPY) \begin{itemize}
\item Data from: 1971-1973
\item Crash Week: 4 October 1998
\item Predicted $|\mathcal{R}|$ for Crash Week: 24.02
\item Actual $|x|$ for Crash Week: 18.90
\item Interval from Pred. to Crash: $\approx$ 15 years
\end{itemize}

\medskip

\noindent Johnson \& Johnson (JNJ) \begin{itemize}
\item Data from: 1980-1982
\item Crash Week: 05 October 2008
\item Predicted $|\mathcal{R}|$ for Crash Week: 25.99
\item Actual $|x|$ for Crash Week: 10.31
\item Interval from Pred. to Crash: $\approx$ 26 years
\end{itemize}

Note that in all cases the data used to approximate the extreme value of a market crash was from years and sometimes decades prior to the crash itself. This implies that each market might have its own unique properties that affect the probability of its price displacement and its price displacement extremes.

We call the property $\hat{m}$ an ``inertial coefficient'' by analogy only, but it appears to act much the same way an object's mass would in the physical world. An asset with a relatively high inertial coefficient is less likely to move as far in price as an asset with a relatively low inertial coefficient. This market property is, at a minimum, consistent with certain behaviors of objects we experience in the physical world.


\section{Conclusions}

Asset prices move either up or down during some period of elapsed time. This up-and-down motion, by definition, is linear for the specified period. Linear motion in one dimension, in this case the dimension of price, is the result we would expect if the asset's mechanics comply with the principle of stationary action. Yet, linear motion in one dimension is also achievable through circular motion in two dimensions. If the single dimension we observe is ``real'' and the unobserved second dimension ``imaginary,'' then circular motion in the complex plane can explain the observable linear motion of assets through the price dimension. 

Given this construct, the phase displacement of the complex wave function and the phase displacement of its complex conjugate are equally likely to produce an observable price displacement. The square of the wave's phase displacement is then responsible for any observable linear motion, and therefore, the stationary action. As a result, we hold that the square of a function of the phase displacement is equal to the action of the asset, or $\mathcal{S}=u^2.$ Applying the action principle to defined periods produces price displacement results for assets that are consistent with historical price data for those periods.

What are some of the implications of this theory and the preliminary results we've seen? The first is that asset price displacements might comply with certain physical laws. We showed here theoretically, and the historical data do not contradict the conclusion, that asset price displacements are perhaps constrained by extreme positive and negative values beyond which whey cannot go for specified intervals of time.

Might this theoretical discovery act as a sort of ``Black Swan'' predictor \cite{taleb}, at least in magnitude? While there is nothing in this theory that would tell us \textit{when} a low-probability event would occur such as the market crash during the week of 5 October 2008, the theory did accurately predict the \textit{magnitude} of the constraint of the price displacement for any trading week, including the first week of the October 2008 crash.

Another way for researchers to test this theory is to look at the correlation between price displacements and the net trading volume for appropriate samples of periods $t.$ If one examines the correlation between the total trading volume and the price displacement for the sample one should find little to no correlation. There should be a sense of randomness. Yet, if one examines the \textit{net} trading volume (i.e., the volume attributed to $\alpha$ less the volume attributed to $\beta$) one should find a fairly strong correlation. Your author's guess is that the strongest correlation will be nonlinear.

Hopefully this theory also inspires a better understanding of market mechanic derivatives, especially asset and options pricing theory. While this paper does not explore these areas it seems only natural that other work specifically examining asset and options pricing might benefit from this different point of view.

Lastly, even though these results seem promising, we should remember that the menu is not the meal and the map is not the terrain. Simply because the results of employing such a model suggest that financial markets might comply with certain physical laws assumed in the model does not mean that this is an exact explanation of phenomena we witness in everyday market mechanics; that is, these very well might not be the ``actual descriptions of the forces and interactions at hand \cite{gallegati}.'' Furthermore, this theory might only explain macro-market movements and might not be as applicable to all individual assets. It's possible that not all individual financial markets have the same mechanical patterns. Again, further research is need to find the limits of this theory's application.

I sincerely appreciate the help, review, and advice of my dear friend Joseph R. Hanley, my father Dr. John T. Manhire, and Professors James McGrath, Lisa Rich, Saurabh Vishnubkat, and Nuno Garoupa at Texas A\&M. I also wish to thank Andrew P. Morriss for our catalytic discussions during the early stages of this work and his continued support thereafter. Most of all, I thank (and apologize to) my wife, Ann, and our nine children who have endured inordinate neglect during the period this problem has consumed their husband and father.
\\[1ex]
\small\textit{College Station, Texas. jmanhire@tamu.edu.}

\end{document}